\documentstyle[prl,aps,twocolumn,psfig]{revtex}
\begin{document}

\twocolumn[\hsize\textwidth\columnwidth\hsize\csname
@twocolumnfalse\endcsname

\draft

\title{Reconstructing the impulse response of a diffusive medium with the
Kramers-Kronig relations}

\author{Er'el Granot and Shmuel Sternklar}

\address{Department of Electrical and Electronics Engineering, College of Judea and Samaria, Ariel 44837, Israel}

\maketitle
\begin{abstract}
\begin{quote}
\parbox{16 cm}{\small

The Kramers-Kronig (KK) algorithm, useful for retrieving the phase
of a spectrum based on the known spectral amplitude, is applied to
reconstruct the impulse response of a diffusive medium. It is
demonstrated by a simulation of a 1D scattering medium with
realistic parameters that its impulse response can be generated
from the KK method with high accuracy. }
\end{quote}
\end{abstract}


]

\narrowtext \footnotetext{erel@yosh.ac.il} \noindent

\section{introduction}
%\emph{I. Introduction.}

Recently there has been a growing interest in light propagation in
diffusive or turbid media \cite{R1}. This can be attributed to
several areas of application-driven research. Diffusive media are
ubiquitous in our environment, and imaging through them is always
a challenge. Clouds, mist, fog, dust and smoke decrease the
visibility on land while surface waves and turbid water reduce
visibility at sea. In the medical field, the main obstacle that a
physician encounters while trying to diagnose internal organs is
the diffusivity rather than the absorption of the biological
tissues. As a consequence, many methods were developed to image
through and to investigate diffusive media \cite{R1}.

One of the most intuitive methods to investigate a diffusive
medium is to measure its' impulse response by a fast detector
\cite{R2,R3,R4,R5,R6}, since, in principle, the impulse response
carries all the optical information of any linear medium. In
practice, fast detectors measure the \textit{intensity} of the
impulse response. The 'first-light' component of the impulse
response carries the ballistic information of the medium. This
information can be used in the reconstruction of the ballistic
image of the diffusive medium. Obviously, since the number of
ballistic photons is extremely small, the image can be partially
reconstructed by the quasi-ballistic (or "snake") photons, which
would impair the spatial resolution of the image.

Assuming that the amount of detected quasi-ballistic photons is sufficient
for the image reconstruction, in order to obtain high resolution images the
detectors must be very fast to distinguish between the (quasi) ballistic and
the diffusive photons. As a consequence such an imaging technique requires
complicated and expensive equipment.

This is one of the main motivations for developing spectral
techniques \cite{R7,R8,R9,R10}, i.e., techniques which work in the
spectral domain instead of the temporal one \cite{R4,R5,R6}. In
one of these techniques the spectral response of the medium
$H\left( \omega \right) = A\left( \omega \right)\exp \left[
{i\varphi \left( \omega \right)} \right]$ is measured for each
wavelength in a wide spectral range \cite{R11,R12,R13}. While an
amplitude $A\left( \omega \right)$ measurement is a relatively
simple task (and fast, since it does not require long averaging),
a phase $\varphi \left( \omega \right)$ measurement is more
complicated since it usually requires interferometric processes
that are inherently complicated and susceptible to fluctuations.
Therefore, phase measurements are of limited value when the medium
varies in time. However, if the phase is reconstructed from the
amplitude measurement and the process is sufficiently fast, the
temporal impulse response can be easily generated even for a
varying medium by a simple inverse Fourier transform. Since in
most cases the diffusive medium can be regarded as linear (and
hopefully time-invariant) system the Kramers-Kronig (KK) method
\cite{R14,R15} can be used to derive the phase from the amplitude
measurement.

Nevertheless, the KK method has several drawbacks. Firstly, in
order to derive the phase one needs the amplitude over the entire
spectrum (from zero to infinity), while experiments can produce
amplitude measurements for only a finite range of frequencies. As
a consequence, any evaluation of phases with the KK method is
always an approximation (see, for example, ref.\cite{R16}). Some
methods were developed to improve these approximations (by
improving the convergence of the integrals) such as the singly and
multiply subtractive KK relations (\cite{R17} and \cite{R18}
respectively).

A more complicated problem arises from the logarithm function,
which diverges whenever the spectral response of the medium
$H\left( \omega \right)$ vanishes. As a consequence the full KK
relation is \cite{R19}

\begin{equation}
\varphi_{KK}(\omega)=-\frac{\omega}{\pi}P\int_{0}^{\infty}d\omega'\frac{\ln|H(\omega')|}{\omega'^2-\omega^2}+\sum_{j}\arg\left(
\frac{\omega-\omega_j}{\omega-\omega^*_j} \right) \label{eq1}
\end{equation}

\noindent
where the $P$ denotes Cauchy's principal value. The first term on the right is
the ordinary KK relation, while the second one is the Blaschke term, which
takes account all of the zeros $\omega _j $ of $H\left( \omega \right)$.

The problem with the zeros of $H\left( \omega \right)$ is more
fundamental since while the first mentioned problem can be
mitigated, at least in principle, by measuring the amplitude for a
larger spectral range, the zeros of $H\left( \omega \right)$
cannot in general be deduced from the amplitude measurements
alone. This problem is confronted mainly in reflection
measurements, where $H\left( \omega \right)$ may vanish at certain
complex frequencies (for a discussion on this subject see refs.
\cite{R20} and \cite{R21}). However, a diffusive medium in the
transmission configuration is a good candidate for the
implementation of the KK method in the following case.

In any diffusive medium the diffusion coefficient of the medium $D$ and its
size $L$ determines the smallest frequency scale $\delta \omega \sim D /
L^2$, which is related to its reciprocal parameter $t_D \sim L^2 / D$ -- the
mean time a photon dwells in the diffusive medium . If the measured spectral
range $\Delta \omega \equiv \omega _{\max } - \omega _{\min } $ is
considerably larger than $\delta \omega $ then the main features of the
impulse response can be retrieved from the KK relations. Moreover, we will
demonstrate that in this regime it is possible to evaluate the amplitude
$\left| {H\left( \omega \right)} \right|$ beyond the spectral range (i.e.,
$\omega < \omega _{\min } $ and $\omega > \omega _{\max } )$ by a certain
average over its values inside $\omega _{\min } < \omega < \omega _{\max }
$. In what follows we will revert to wavenumbers instead of frequencies, but
the transition between the two is trivial.

In most previous works the desired parameter was the refractive
index of a medium, so that the KK method was mostly implemented
for cases where the attenuation was caused by absorption rather
than by scattering. As a result, using the KK method as a tool to
measure the impulse response of a diffusive medium was not common.
Recently, we demonstrated \cite{R12} that the KK method, even in
its simplest and most naive form can be used to reconstruct the
impulse response of a Fabry-Perot interferometer, whose response
is governed solely by scattering. It was shown that with
relatively simple equipment its' impulse response can be evaluated
with very high temporal resolution (less than 200fs).

In this paper we demonstrate by a numerical simulation that the KK method
can be useful in determining the phases of the transfer function of a
diffusive medium. It is also shown that the calculated spectral response can
be used in reconstructing the medium's impulse response with high accuracy.

\section{The model}

We investigate a 1D homogenous medium with $N$ small scatterers,
each having a different refractive index and width, randomly
distributed in the medium. For simplicity it is assumed that the
width of each scatterer $l_j $ is considerably smaller than the
beam's wavelength $\lambda $ (i.e., $l_j < < \lambda $ for every
$j)$, however, this is not a restrictive assumption since a
diffusive medium, whose dimensions are considerably larger than
the medium's diffusion length is characterized, almost by
definition, only by a median value for the scattering coefficient
(i.e., as long as the set of scattering coefficients are similar
the fine structure of each scatterer is not important; similarly,
the exact locations of the scatterers have a negligible effect on
the diffusion coefficient).

For simplicity we ignore polarization effects, and thus the 1D
stationary-state wave equation has the form

\begin{equation}
\label{eq2} \frac{\partial ^2}{\partial x^2}\psi ^2 + \left(
{\frac{2\pi n\left( x \right)}{\lambda }} \right)^2\psi = 0
\end{equation}

\noindent
where $\psi \left( x \right)$ is the electromagnetic field and $n\left( x
\right)$ is the refractive index.

In general, the refractive index of a diffusive medium, modeled as a
homogenous medium with multiple random scatterers, has the general form
presented in fig.1. For simplicity, however, we choose to simulate the
variations in the refractive index of the medium by delta functions, which
corresponds to scatterers whose dimensions $l_j $ are considerably smaller
than the beam's wavelength $l_j < < \lambda $, consistent with our previous
assumption. It should be stressed, that any small 1D scatterer (with respect
to the wavelength) can be replaced for any practical reason with a delta
function, the reason being that neither its width $l_j $ nor its strength
$\delta n_j $ (the difference between its refraction index and the
surrounding) appears separately in the scattering solutions, only their
product $l_j \delta n_j $ does. Therefore, a small scatterer, whose width
and strength are $l_j $ and $\delta n_j $ respectively can be replaced by a
delta function whose prefactor is proportional to $l_j \delta n_j $ (see
below for details).

\begin{figure}
\psfig{figure=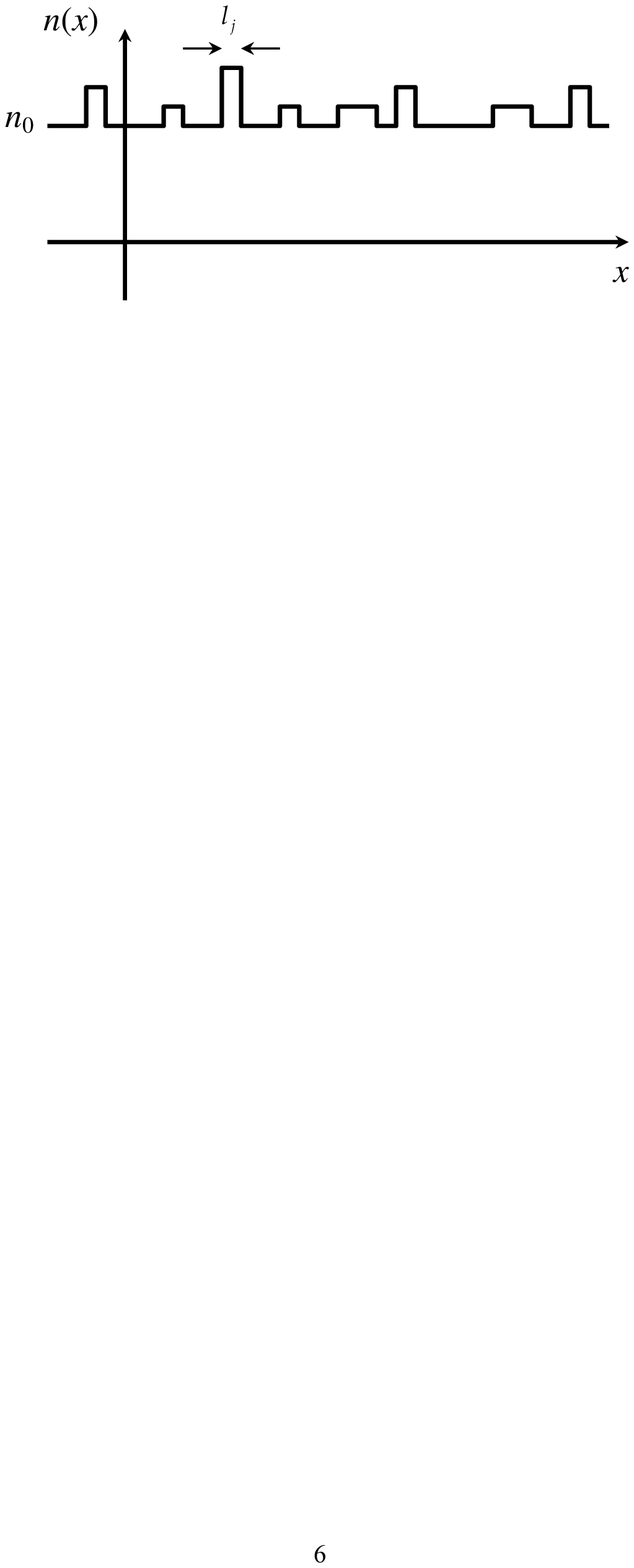,width=8cm,bbllx=130bp,bblly=600bp,bburx=455bp,bbury=780bp,clip=}
\caption{\emph{Refractive index as a function of location for a
diffusive medium }}\label{fig1}
\end{figure}

The \textit{square} of the refraction index can be separated into homogenous ($n_0^2 )$ and
varying ($2n_0 \Delta n)$ parts

\begin{equation}
\label{eq3} \frac{\partial ^2}{\partial x^2}\psi ^2 + \left(
{\frac{2\pi }{\lambda }} \right)^2\left[ {n_0^2 + 2n_0 \Delta
n\left( x \right)} \right]\psi = 0
\end{equation}

With the definition of the wavenumber $k \equiv 2\pi n_0 / \lambda $ the
wave equation can be written

\begin{equation}
\label{eq4} \frac{\partial ^2}{\partial x^2}\psi + k^2\left( {1 +
2\frac{\Delta n\left( x \right)}{n_0 }} \right)\psi = 0,
\end{equation}

\noindent
where $2\Delta n\left( x \right) / n_0 = \sum\limits_{j = 1}^N {\alpha _j
\delta \left( {x - L_j } \right)} $, $N$ is the number of scatterers, $L_j =
L_{j - 1} + a_j = \sum\limits_{m = 1}^j {a_m } $ is the position of the
$j$th scatterer, i.e., $a_j $ are the distances between two adjacent
scatterers, and

\begin{equation}
\label{eq5} \alpha _j = 2\frac{\delta n_j l_j }{n_0 }
\end{equation}

\noindent
is the strength of the $j$th scatterer, where $\delta n_j $ and $l_j $ are the
change in its refractive index and its size respectively.

Therefore, the wave equation for the diffusive medium reads

\begin{equation}
\label{eq6} \frac{\partial ^2}{\partial x^2}\psi + k^2\left[ {1 +
\sum\limits_{j = 1}^N {\alpha _j \delta \left( {x - L_j } \right)}
} \right]\psi = 0 \quad .
\end{equation}

\begin{figure}
\psfig{figure=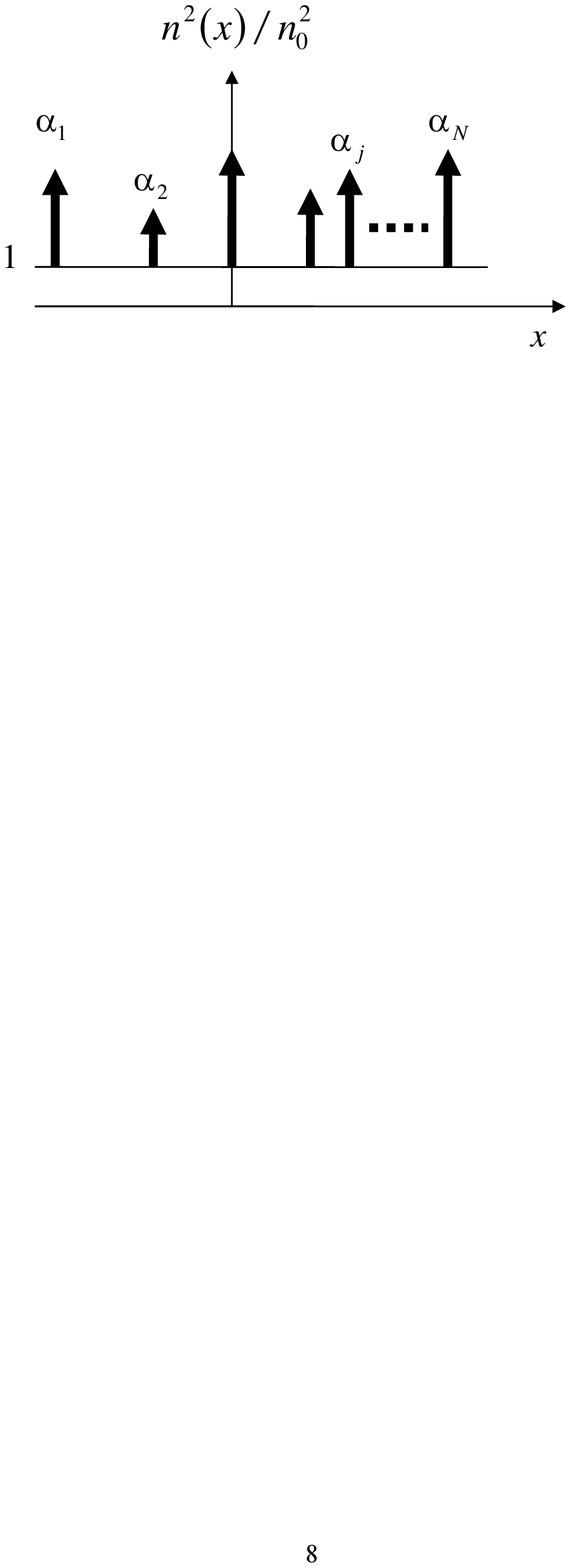,width=8cm,bbllx=135bp,bblly=570bp,bburx=435bp,bbury=770bp,clip=}
\caption{\emph{A schematic presentation of the diffusive medium.
Each scatterer is presented in the model as a delta function
change in the refractive index. }}\label{fig2}
\end{figure}

In the appendix we elaborate on the derivation of the medium's
spectral response $H\left( k \right)$ from eq.\ref{eq6}. We then
apply the KK method to a finite spectrum $k_{\min } \le k \le
k_{\max } $ to determine the phase:

\begin{equation}
\label{eq7} \varphi _{KK} \left( k \right) = - \frac{k}{\pi
}P\int\limits_{k_{\min } }^{k_{\max } } {dk'\frac{\ln \left|
{H\left( {k'} \right)} \right|}{k'^2 - k^2}} .
\end{equation}

\noindent
which is then substituted into:

\begin{equation}
\label{eq8} H_{KK} \left( k \right) = \left| {H\left( k \right)}
\right|\exp \left[ {i\varphi _{KK} \left( k \right)} \right]
\end{equation}

\noindent to evaluate the medium's spectral response. If we keep
$\Delta k \equiv k_{\max } - k_{\min } > > Dn / \left( {cL^2}
\right)$ (where $n$ is the refractive index of the medium) then
most of the features of the impulse response can be derived.
Moreover, in this regime, approximation (\ref{eq7}) can be
improved by noticing that the \textit{mean} value of $\ln \left|
{H\left( k \right)} \right|$ does not change much beyond the
measured region $k_{\min } \le k \le k_{\max } $ (at least in its
spectral vicinity). Therefore, the boundaries of the integral
(\ref{eq7}) can be taken as 0 and $\infty $ while $\ln \left|
{H\left( k \right)} \right|$ for $0 < k < k_{\max } $and $k_{\max
} < k < \infty $ is replaced with its average value, i.e.,

\begin{equation}
\label{eq9} \begin{array}{l}
 \varphi _{KK} \left( k \right) = - \frac{k}{\pi }P\int\limits_{k_{\min }
}^{k_{\max } } {dk'\frac{\ln \left| {H\left( {k'} \right)}
\right|}{k'^2 -
k^2}} - \\
 \frac{k}{\pi }\left\langle {\ln \left| {H\left( k \right)} \right|}
\right\rangle \left[ {\int\limits_0^{k_{\min } } {dk'\frac{1}{k'^2
- k^2} +
\int\limits_{k_{\max } }^\infty {dk'\frac{1}{k'^2 - k^2}} } } \right] \\
 \end{array}
\end{equation}

\noindent
where $\left\langle {\ln \left| {H\left( k \right)} \right|} \right\rangle
\equiv \left( {k_{\max } - k_{\min } } \right)^{ - 1}\int_{k_{\min }
}^{k_{\max } } {dk'} \ln \left| {H\left( {k'} \right)} \right|$ is the mean
value of $\ln \left| {H\left( k \right)} \right|$ in the measured region.

It should be noted that in cases where the KK method is used to
derive the refractive index (as is the case in negligibly
scattering media) the variations in $\ln \left| {H\left( k
\right)} \right|$ are of the same scale as the spectral range
$\Delta k$, i.e., $d\ln \left| {H\left( k \right)} \right| /
dk\sim \Delta k^{ - 1}$. Therefore, in these cases analytical
continuation and extrapolations are used to approximate $\ln
\left| {H\left( k \right)} \right|$ beyond the measured region
\cite{R22}. In the scattering medium case the situation is
considerably different, namely $d\ln \left| {H\left( k \right)}
\right| / dk >
> \Delta k^{ - 1}$, and therefore extrapolations has little value.
On the other hand the spectral variations are so strong that they
rapidly converge to the average value.

By solving the integrals one obtains an evaluation of the phases from the
amplitude measurements

\begin{equation}
\label{eq10} \varphi _{KK} \left( k \right) = - \frac{k}{\pi
}P\int\limits_{k_{\min } }^{k_{\max } } {dk'\frac{\ln \left|
{H\left( {k'} \right)} \right|}{k'^2 - k^2}} + \Delta \varphi
_{KK} \left( k \right)
\end{equation}

\noindent
where

\begin{equation}
\label{eq11} \Delta \varphi _{KK} \left( k \right) \equiv -
\frac{1}{2\pi }\left\langle {\ln \left| {H\left( k \right)}
\right|} \right\rangle \ln \left( {\frac{k - k_{\min } }{k +
k_{\min } } \cdot \frac{k_{\max } + k}{k_{\max } - k}} \right)
\end{equation}

\noindent
is a correction term.

This phase is then substituted into eq. (\ref{eq8}) to reconstruct
the full transfer function.

If the spectrum of the input signal is a rectangular function in
the spectral domain $k_{\min } \le k \le k_{\max } $ (and
therefore the field is proportional to $E_{in} \left( t \right)
\propto \exp \left( {i\bar {k}ct} \right){\mathrm{sinc}}\left(
{\Delta kct / 2} \right)$, where $\Delta k \equiv k_{\max } -
k_{\min } $ and $\bar {k} \equiv \left( {k_{\max } + k_{\min } }
\right) / 2)$ the exact impulse response, which will be measured
at the end of the medium, i.e., at $x > L_N $, is

\begin{equation}
\label{eq12} I\left( t \right) = \left| {\int\limits_{k_{\min }
}^{k_{\max } } {dk'H\left( {k'} \right)\exp \left( {ik'ct}
\right)dk'} } \right|^2
\end{equation}

\noindent
while the KK reconstruction, which is based only on the amplitude (or
intensity) measurements, predicts

\begin{equation}
\label{eq13} I_{KK} \left( t \right) = \left|
{\int\limits_{k_{\min } }^{k_{\max } } {dk'H_{KK} \left( {k'}
\right)\exp \left( {ik'ct} \right)dk'} } \right|^2.
\end{equation}

We will demonstrate by a simulation that in a realistic case the latter
expression is a very good approximation to the former.

\section{Simulation}

For simplicity we choose air as the homogenous medium, i.e., $n_0
= 1$ (however, the results can easily be scaled to any dielectric
medium with an arbitrary $n_0 )$ with 150 small scatterers, where
the distance between them is a random variable, distributed
uniformly between 0 and $5\mu m$ (i.e., $0 \le a_j \le 5\mu m)$.
Similarly, the strength of each scatterer is also a uniform random
number so that $0 \le \alpha _j \le 0.03\mu m$ (this corresponds
to glass particles having widths on the order of tens of
nanometers).

\begin{figure}
\psfig{figure=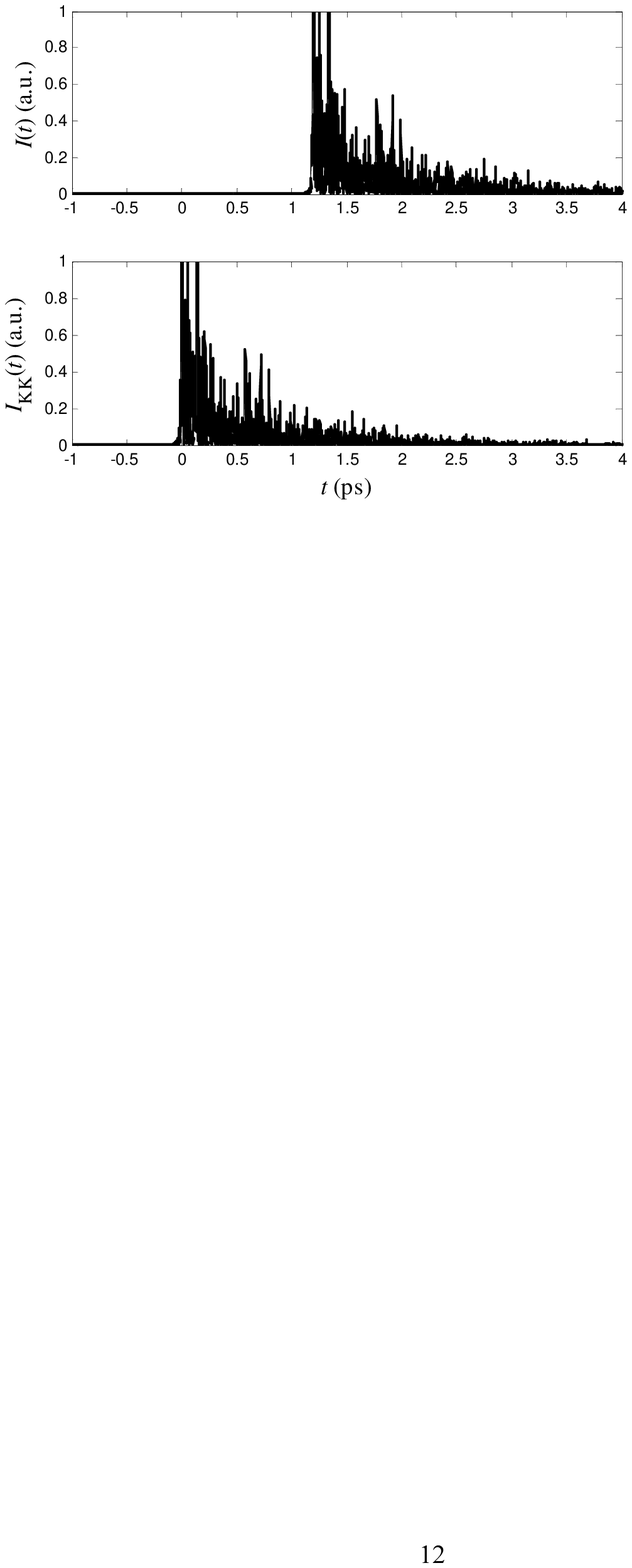,width=8cm,bbllx=80bp,bblly=510bp,bburx=420bp,bbury=770bp,clip=}
\caption{\emph{The impulse response of the medium. The upper plot
is the exact reconstruction, while the lower one is the KK
reconstruction. }}\label{fig3}
\end{figure}

We assume that the incoming pulse that penetrates the medium has a
rectangular spectral shape, i.e., $E_{in} \left( t \right) \propto
\exp \left( {i\bar {k}ct} \right){\mathrm{sinc}}\left( {\Delta kct
/ 2} \right)$, with $\bar {\omega } \equiv \bar {k}c = 2\pi \times
750\mbox{THz}$ and $ \Delta \omega \equiv \Delta kc = 2\pi \times
300\mbox{THz}$.

Since $D \cong cl_{RW} / n$ where $l_{RW} \cong a / R$ ($a$ is the average
distance between scatterers and $R$ is the mean reflection coefficient) is
the random walk in the diffusion process, then the minimum spectral range
required is $\delta k \cong a / RL^2 = \left( {NRL} \right)^{ - 1}$ (where
$N$ is the total number of scatterers).

In this case the measured spectral range is six orders of magnitude larger
than $\delta k$ i.e., $\delta k \cong 15m^{ - 1} < < 6.3\times 10^6m^{ - 1}
\cong \Delta k$.

In Fig.3 the reconstruction of the impulse responses of the exact
solution $I\left( t \right)$ and the KK reconstruction $I_{KK}
\left( t \right)$ are presented.

\begin{figure}
\psfig{figure=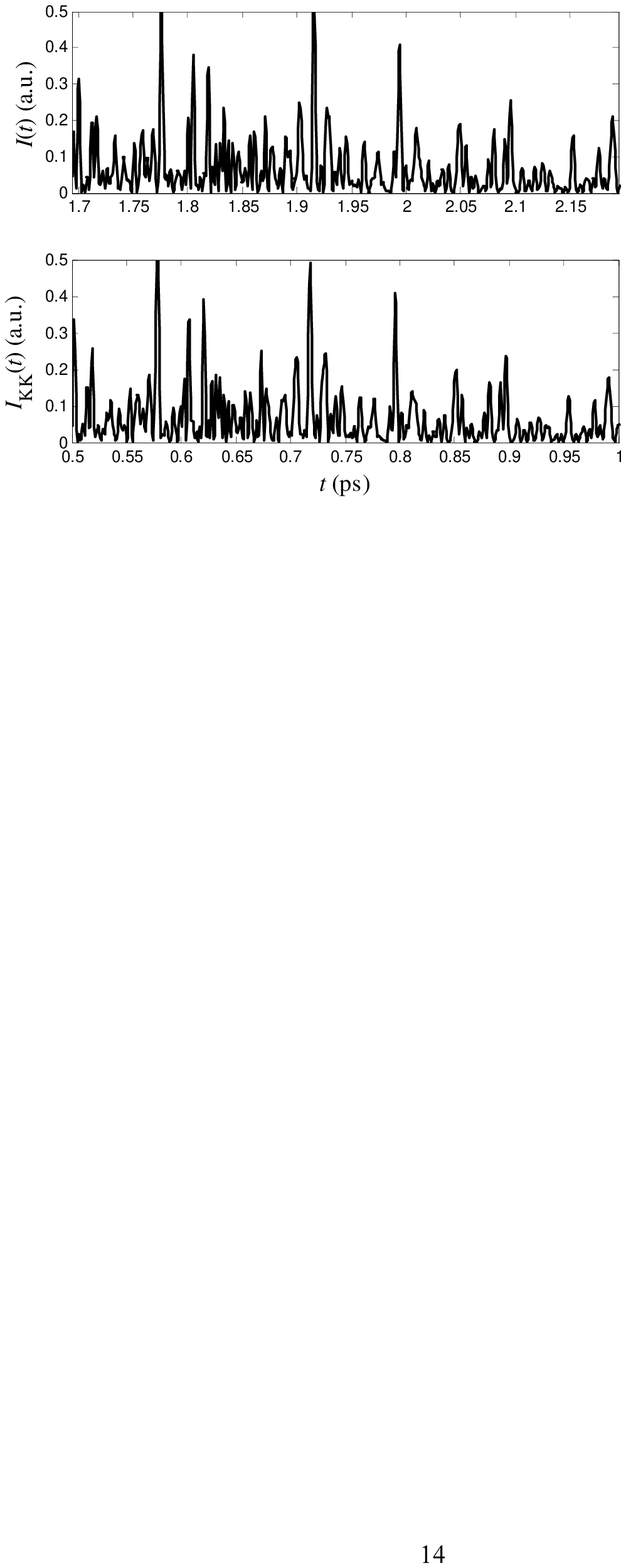,width=8cm,bbllx=80bp,bblly=520bp,bburx=410bp,bbury=780bp,clip=}
\caption{\emph{Zoom-in of fig.3 }}\label{fig4}
\end{figure}

As can be seen, except for the delay between the two pulses, which
is a direct consequence of the KK technique, the two signals are
very similar. Evidently, they are not identical but the
differences between them are quite small. In Fig.4 we reveal
details of the temporal response and compare the two responses
over a small temporal window, showing that the two are alike even
on the microscopic level.

\begin{figure}
\psfig{figure=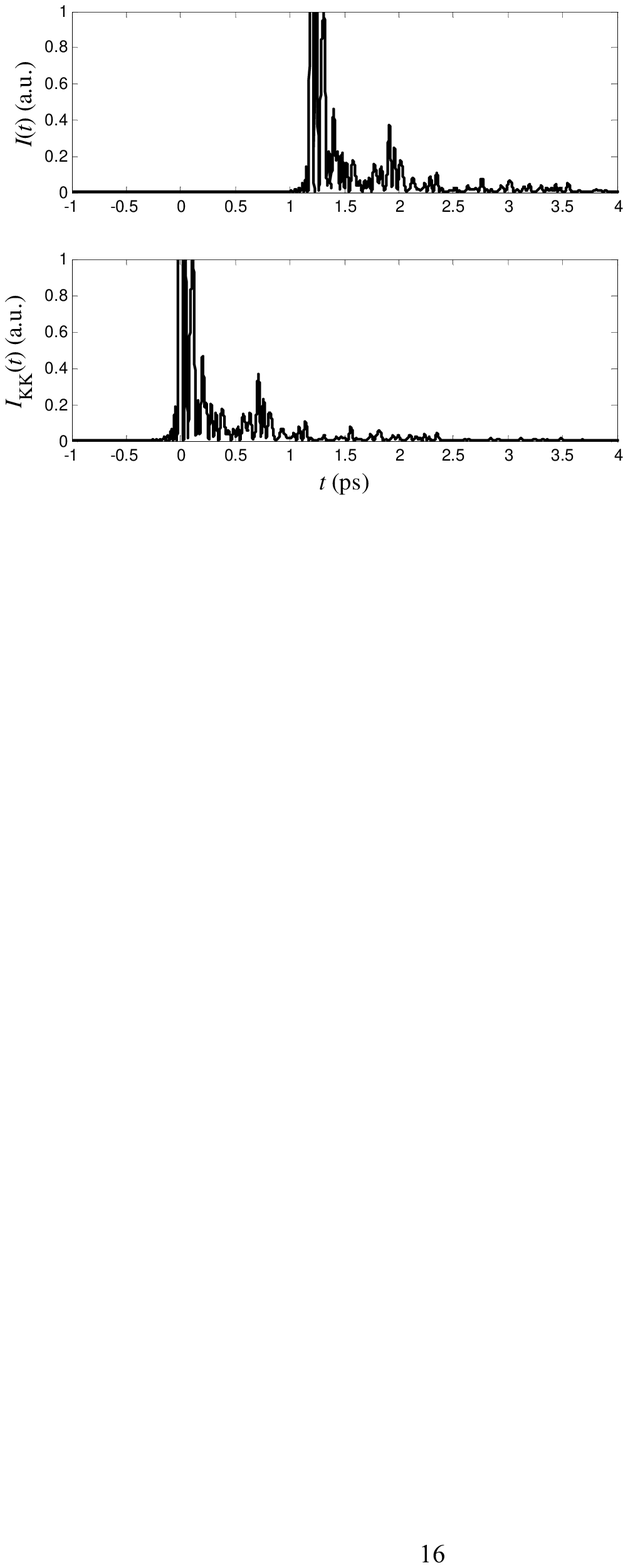,width=8cm,bbllx=80bp,bblly=520bp,bburx=410bp,bbury=780bp,clip=}
\caption{\emph{Same as Fig.3 but with a spectral width of $\Delta
\omega = 2\pi \times 30\mbox{THz}$. }}\label{fig5}
\end{figure}

If the spectrum of the incoming pulse is narrower, i.e., the
incoming pulse is temporally broader, the spikes in the outgoing
pulse are respectively wider. In Fig. 5 and 6 the impulse response
of the same system is presented when the spectral width of the
incoming pulse is ten times narrower (than in Figs. 3 and 4), that
is, $\Delta \omega = 2\pi \times 30\mbox{THz}$. As can be seen
from these plots, the KK approximation is even better for the
spectrally narrow pulse. Therefore, although the integration in
the KK expressions covers a narrower region, and in principle
should yield inferior results, since the spikes are temporally
wider and therefore are less susceptible to dispersion
deformation, the reconstruction is improved. The problem is,
however, that as $\Delta k$ decreases and gets closer to $\delta
k$ there is insufficient spectral information to reconstruct the
complete impulse response and the error in the evaluation of
macroscopic averages (such as the diffusion coefficient)
increases.

\begin{figure}
\psfig{figure=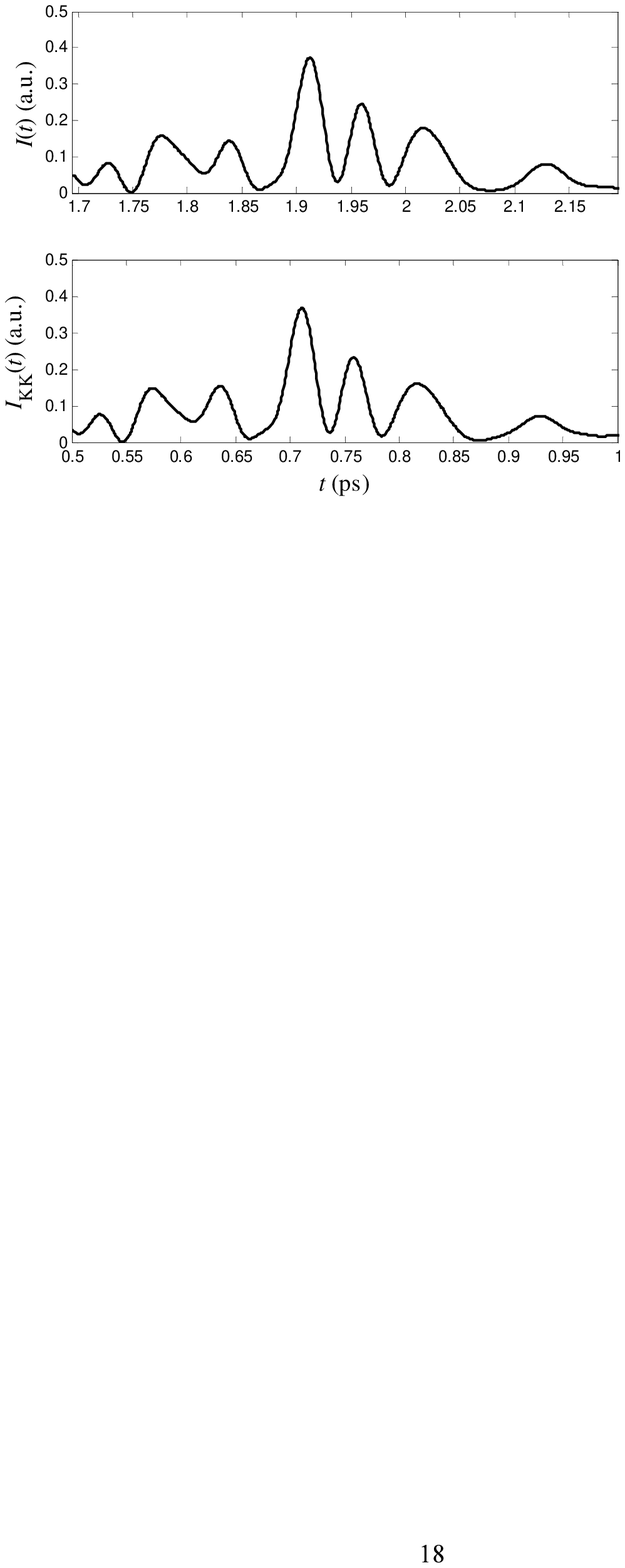,width=8cm,bbllx=80bp,bblly=520bp,bburx=410bp,bbury=780bp,clip=}
\caption{\emph{Same as Fig.4 but with a spectral width of $\Delta
\omega = 2\pi \times 30\mbox{THz}$. }}\label{fig6}
\end{figure}

\section{The effect of the correction term}

To illustrate the importance of the correction term $\Delta
\varphi _{KK} \left( k \right)$ in eq.\ref{eq8} we repeat the
simulation for a bandwidth $\Delta \omega \equiv \Delta kc = 2\pi
\times 150\mbox{THz}$ with and without the CT.

In the upper panel of Fig.7 the reconstruction was made with the CT, and in
the lower panel the CT was absent. There is a clear improvement in the upper
panel.

\begin{figure}
\psfig{figure=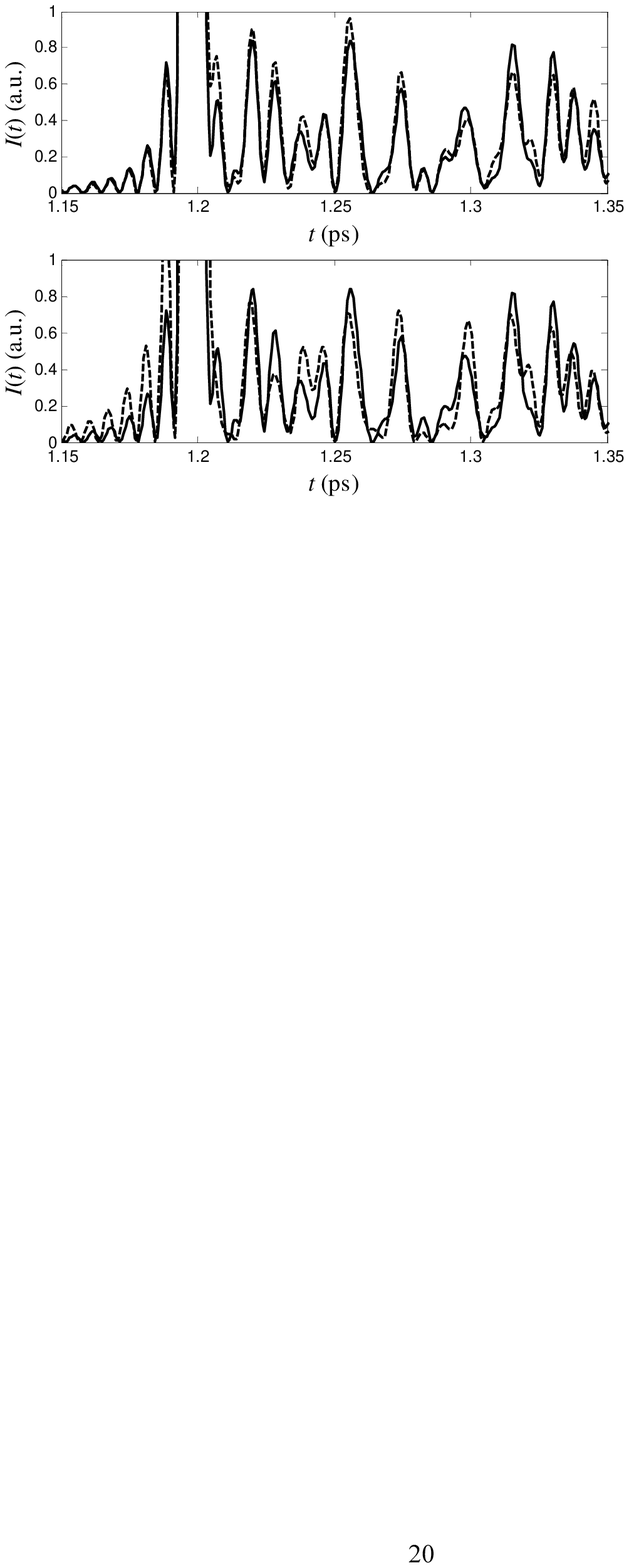,width=8cm,bbllx=80bp,bblly=520bp,bburx=410bp,bbury=780bp,clip=}
\caption{\emph{In upper figure the reconstruction was done with
the CT, while in lower one the calculations were done without it.
In both cases the solid line represents the direct calculation,
and the dashed line corresponds to the KK reconstruction.
}}\label{fig7}
\end{figure}

To understand the influence of the correction term we can expand
it with respect to the zero-correction point $k_0 \equiv \sqrt
{k_{\min } k_{\max } } $

\begin{equation}
\label{eq14} \begin{array}{l}
 \Delta \varphi _{KK} \left( k \right) = \\
 - \frac{1}{2\pi }\left\langle {\ln \left| {H\left( k \right)} \right|}
\right\rangle \left[ {4\kappa - 2\eta \kappa ^2 +
\frac{16}{3}\left( {1 + \frac{8}{3}\eta ^2} \right)\kappa ^3 +
O\left( {\kappa ^4} \right)} \right]
\\
 \end{array}
\end{equation}

\noindent
where $\eta \equiv \frac{\Delta k}{k_0 }$ is a relatively small parameter,
which characterizes the normalized spectral width ($\Delta k \equiv k_{\max
} - k_{\min } )$ and $\kappa \equiv \frac{k - k_0 }{\Delta k}$ is the
normalized wavenumber.

The first term in the expansion is responsible for a constant time delay,
which therefore has a trivial influence on the solution. Moreover, this
linear delay, unless $\left| {H\left( k \right)} \right|$ is very small due
to the multiple scattering, is on the same scale as the initial pulse width
$\tau \equiv \left( {c\Delta k} \right)^{ - 1}$,

\begin{equation}
\label{eq15} \tau _{delay}^{linear} \cong d\Delta \varphi
_{KK}^{linear} \left( k \right) / d(ck) = - \frac{2}{\pi
}\left\langle {\ln \left| {H\left( k \right)} \right|}
\right\rangle \tau .
\end{equation}

When $\Delta k < < k_0 $, i.e., $\eta < < 1$ the second term, which is the
first dispersion term, can be neglected. Therefore, the first term which
causes the main deformation is of the third order, so that

\begin{equation}
\label{eq16} \Delta \varphi _{KK}^{non - linear} \left( k \right)
\cong - \frac{8}{3\pi }\left\langle {\ln \left| {H\left( k
\right)} \right|} \right\rangle \kappa ^3.
\end{equation}

The dispersion coefficient is therefore proportional to
$\left\langle {\ln \left| {H\left( k \right)} \right|}
\right\rangle $.

Since $\left| \kappa \right| \le 1$ the time-delay, which is a consequence
of this term is bounded by

\begin{equation}
\tau _{delay}^{non - linear} \cong \frac{d\Delta \varphi
_{KK}^{non - linear} \left( k \right)}{d(ck)} < - \frac{2}{\pi
}\left\langle {\ln \left| {H\left( k \right)} \right|}
\right\rangle \tau ,
\end{equation}

which is independent of the spectral width $\Delta k$. From this
respect, the deformation in the signal is proportional to the
peaks' temporal width, however, when $\Delta k$ decreases the
average $\left\langle {\ln \left| {H\left( k \right)} \right|}
\right\rangle $ is a better approximation to the value of $\ln
\left| {H\left( k \right)} \right|$ outside the measured spectral
width $\Delta k$, and therefore the KK reconstruction is improved.

\section{Summary}

We have demonstrated through numerical simulations that the KK method can be
used to reconstruct the impulse response of a scattering medium with
realistic parameters. It was shown that when the measured spectral width is
considerably larger than the reciprocal of the diffusion length, i.e.
$\Delta k \equiv k_{\max } - k_{\min } > > Dn / \left( {cL^2} \right)$, the
KK method yields a satisfactory prediction of the impulse response.
Moreover, it was demonstrated that it is possible to take advantage of the
fact that in a diffusive medium the spectral variations are very large but
their running average has relatively small variations. Therefore, the
integrand of the KK relations can be evaluated even outside the measured
spectral domain as the average value of the measurements. It was shown that
this evaluation improves the reconstruction of the impulse response.

Although we focused in this work on the \textit{optical} properties of a diffusive medium,
this technique can be implemented to \textit{any} wave dynamics in diffusive media,
e.g., acoustic scattering (photon scattering), x-ray scattering, electron
scattering etc.

Owing to the simplicity and speed of the KK method, we believe that this
technique is a promising tool for characterizing and imaging through
diffusive media.

\bigskip

This research was supported by the ISRAEL SCIENCE FOUNDATION (grant no.
144/03-11.6).

\section{Appendix A: Calculations of the spectral response
(transfer) function $H\left( k \right)$}

\begin{figure}
\psfig{figure=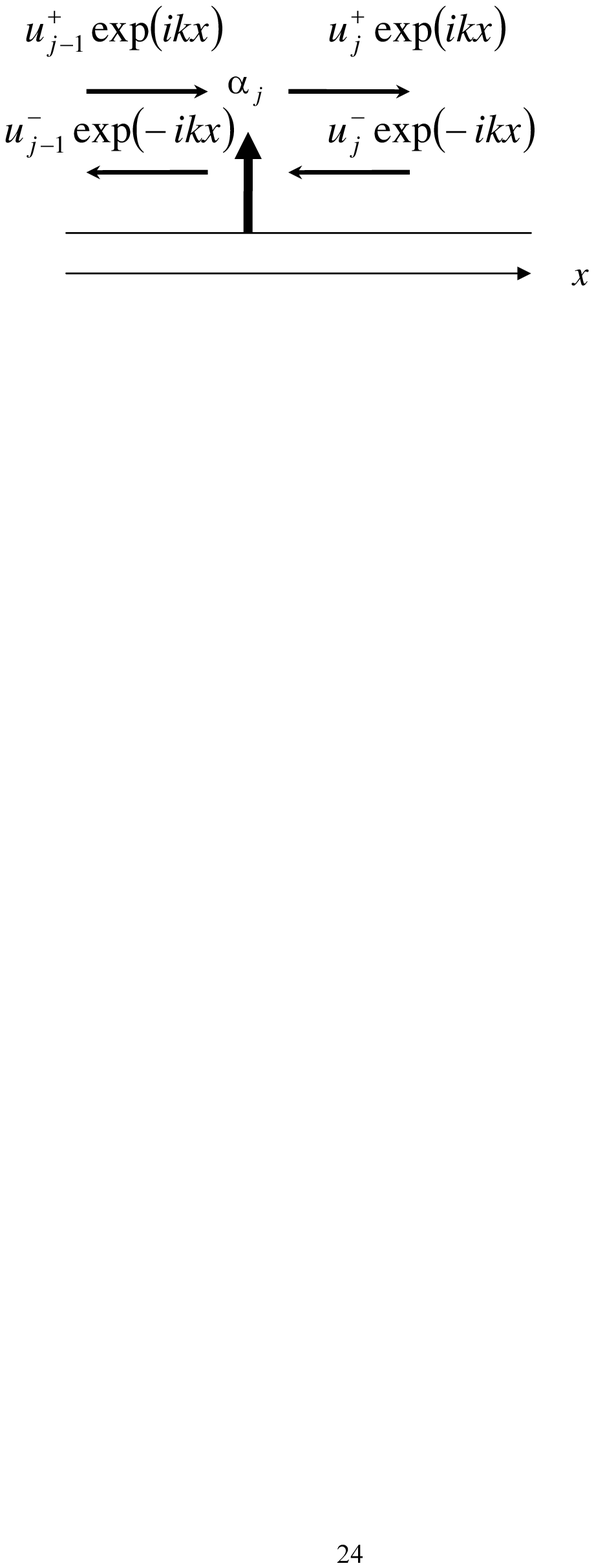,width=8cm,bbllx=110bp,bblly=580bp,bburx=430bp,bbury=750bp,clip=}
\caption{\emph{An illustration of the incoming and outgoing waves
at the vicinity of a single scatterer. }}\label{figA1}
\end{figure}

Since we are discussing a 1D medium, consisting of the scatterers
in an otherwise totally homogenous environment, then in every
point in space between two scatterers the field of the incoming
and outgoing waves (see Fig.8) can be described by two
coefficients. Therefore, for a given $k$ the vector

\begin{equation}
\label{eqA1} {\rm {\bf v}}_j = \left( {{\begin{array}{*{20}c}
 {u_j^ + } \hfill \\
 {u_j^ - } \hfill \\
\end{array} }} \right)
\end{equation}

\noindent
fully describes the field between the $j$th and ($j+$1)th scatterers, which can be
written as

\begin{equation}
\psi \left( {x_j < x < x_{j + 1} } \right) = u_j^ + \exp \left(
{ikx} \right) + u_j^ - \exp \left( { - ikx} \right). \label{eqA2}
\end{equation}

By applying the continuity of the field at the two ends of the
scatterer at $x = 0$

\begin{equation}
\label{eqA3} \psi \left( {x = - 0} \right) = \psi \left( {x = + 0}
\right),
\end{equation}

\noindent
which can be written as
\begin{equation}
\label{eqA4} u_j^+ + u_j^- = u_{j + 1}^ + + u_{j + 1}^ - .
\end{equation}

Similarly, the discontinuity of the field at $x = 0$

\begin{equation}
\label{eqA5} \left. {\frac{\partial }{\partial x}\psi } \right|_{x
= + 0} - \left. {\frac{\partial }{\partial x}\psi } \right|_{x = -
0} = - \alpha _j \psi \left( 0 \right)
\end{equation}

\noindent
can be written

\begin{equation}
\label{eqA6} ik\left( {u_{j + 1}^ + - u_{j + 1}^ - - u_j^ + + u_j^
- } \right) = - \alpha \left( {u_j^ + + u_j^ - } \right)
\end{equation}

Therefore, the influence of each of the scatterers can be describe by the
2x2 matrix,

\noindent
which relates ${\rm {\bf v}}_j = \left( {{\begin{array}{*{20}c}
 {u_j^ + } \hfill \\
 {u_j^ - } \hfill \\
\end{array} }} \right)$ to ${\rm {\bf v}}_{j + 1} = \left(
{{\begin{array}{*{20}c}
 {u_{j + 1}^ + } \hfill \\
 {u_{j + 1}^ - } \hfill \\
\end{array} }} \right)$ by

\begin{equation}
\label{eqA7} {\rm {\bf v}}_{j + 1} = A_j {\rm {\bf v}}_j
\end{equation}

\noindent
where

\begin{equation}
\label{eqA8} A_j = \left( {{\begin{array}{*{20}c}
 {1 - i\eta _j } \hfill & {i\eta _j } \hfill \\
 { - i\eta _j } \hfill & {1 + i\eta _j } \hfill \\
\end{array} }} \right)
\end{equation}

\noindent
and $\eta _j \equiv \alpha _j k / 2$. The homogenous medium between the
$j$th and ($j+$1)th scatterers can be described by the matrix

\begin{equation}
\label{eqA9} D_j = \left( {{\begin{array}{*{20}c}
 {\exp \left( {ika_j } \right)} \hfill & 0 \hfill \\
 0 \hfill & {\exp \left( { - ika_j } \right)} \hfill \\
\end{array} }} \right).
\end{equation}

With these two types of matrices, one can generate the matrix of the medium
that includes the first $j$ scatterers $M_j $ if the matrix of the $j$-1
scatterers $M_{j - 1} $ is known

\begin{equation}
\label{eqA10} M_j = A_j D_j M_{j - 1} .
\end{equation}

Therefore, with a simple recursion the matrix of a medium $M_N $ with an
arbitrary number of scatteres $N$ can easily be generated.

Finally, the transfer function $H\left( k \right)$, which in our case is the
transmission coefficient of the medium, can easily be reconstructed from the
matrix of the entire medium

\begin{equation}
\label{eqA11}
M_N = \left( {{\begin{array}{*{20}c}
 {m_{11} } \hfill & {m_{12} } \hfill \\
 {m_{21} } \hfill & {m_{22} } \hfill \\
\end{array} }} \right)
\end{equation}

\noindent
by

\begin{equation}
\label{eqA12} H\left( k \right) = \det \left( {M_N } \right) /
m_{22}
\end{equation}

\noindent
for every $k$.

\end{document}